%% file: main.tex
\RequirePackage{fix-cm}
\documentclass[smallextended]{svjour3}       %
\smartqed  %
\usepackage{graphicx}
\usepackage{soul}
\usepackage{dblfloatfix}
\usepackage{arydshln}
\usepackage{url}
\usepackage{booktabs}
\usepackage{cite}
\usepackage{amsmath,amssymb,amsfonts}
\begin{document}

\title{Reinforcement Learning for Online Testing of Autonomous Driving Systems: a Replication and Extension Study}

\author{Luca Giamattei \and
        Matteo Biagiola \and Roberto Pietrantuono \and Stefano Russo \and Paolo Tonella%
}

\institute{Luca Giamattei* [0000-0003-3767-4036], Roberto Pietrantuono [0000-0003-2449-1724], Stefano Russo [0000-0002-8747-3446] \at
              Università di Napoli Federico II\\
              Via Claudio 21, 80125 -- Napoli, Italy \\
                Phone: +39 0817683820 -- Fax: +39 0817683816\\
              \email{\{luca.giamattei, roberto.pietrantuono, stefano.russo\}@unina.it}\\
              \textit{*corresponding author}%
           \and
           Matteo Biagiola, Paolo Tonella [0000-0003-3088-0339] \at
              Università della Svizzera Italiana\\
              Via Buffi, 13 -- Lugano, Switzerland \\
            Phone: +41 58 666 40 00 -- Fax: +41 58 666 46 47 \\
            \email{\{matteo.biagiola, paolo.tonella\}@usi.ch}
}

\date{Received: date / Accepted: date}

\maketitle

\begin{abstract}
In a recent study, Reinforcement Learning (RL) used in combination with many-objective search, has been shown to outperform alternative techniques (random search and many-objective search) for online testing of Deep Neural Network-enabled systems. 
The empirical evaluation of these techniques was conducted on a state-of-the-art Autonomous Driving System (ADS). 
This work is a replication and extension of that empirical study. 
Our replication shows that RL does not outperform pure random test generation in a comparison conducted under the same settings of the original study, but with no confounding factor coming from the way collisions are measured. 
Our extension aims at eliminating some of the possible reasons for the poor performance of RL observed in our replication: (1) the presence of reward components providing contrasting or useless feedback to the RL agent; (2) the usage of an RL algorithm (Q-learning) which requires discretization of an intrinsically continuous state space.
Results show that our new RL agent is able to converge to an effective policy that outperforms random testing.
Results also highlight other possible improvements, which open to further investigations on how to best leverage RL for online ADS testing.

\end{abstract}

\section{Introduction}
\label{sec:introduction}
\input{sections/introduction.tex}

\section{Background} 
\label{sec:background}
\input{sections/background.tex}

\section{Replication study}
\label{sec:replicated_study}

\input{sections/replicated_study.tex}

\section{Extension study}
\label{sec:extension_study}
\input{sections/extension_study.tex}

\section{Related work} 
\label{sec:related}
\input{sections/related.tex}

\section{Conclusions}
\label{sec:conclusions}
\input{sections/conclusions.tex}

\section*{Data Availability}
For the sake of Open Science, we provide all artifacts for replication and the results of our studies.
Available at: \url{https://doi.org/10.6084/m9.figshare.24794544}.

\section*{Acknowledgements}
This work was partially supported by the H2020 project PRECRIME, funded under the ERC Advanced Grant 2017 Program (ERC Grant Agreement n. 787703).
It has also received funding from the European Union’s Horizon 2020 research and innovation programme under the Marie Sk{\l}odowska-Curie grant agreement No 871342 “uDEVOPS”. 
This material is supported by the Google Research Credits program with the award GCP291784478.

\end{document}

%% file: sections/introduction.tex
Testing of Deep Neural Network-enabled systems is a challenging and expensive task - yet essential - in the engineering of many modern systems with artificial intelligence components at their core.
Testing of Autonomous Driving Systems (ADSs) is gaining particular attention in the scientific community, as any advancement towards more effective and efficient techniques increases safety and reduces costs.

In the literature, a common way to test ADSs is to manipulate the simulation environment in which the ADS under test operates~\cite{lei-ma-survey}. The objective of such techniques is to perturb the environment to try to cause a misbehavior of the ADS, e.g., a collision with another vehicle %
or a traffic rule violation. Tests are generated automatically by solving an optimization problem, which consists of finding the optimal configurations of the objects in the environment to maximize an objective function (usually the distance of the ADS from misbehavior). Researchers have proposed search-based techniques to address this optimization problem~\cite{lei-ma-survey,deepjanus,deephyperion,Li20,asfault,icse18-abdessalem}, which have shown to be effective at generating static configurations of the environment that challenge the ADS under test.

However, search-based techniques struggle to deal with sequential interactions at runtime, which are required to manipulate dynamic objects in the environment (e.g., another vehicle). On the other hand, the Reinforcement Learning (RL) paradigm requires the agent to dynamically interact with the environment, learning from the effects of its actions. This paradigm offers an alternative way to test ADSs, by formulating the testing problem as an RL problem. Afterward, the testing technique needs to choose a suitable RL algorithm to learn the appropriate actions that maximize the reward.

In a recent paper presented at the International Conference on Software Engineering (ICSE) in 2023, Haq \textit{et al.}~\cite{Haq23} presented MORLOT (Many-Objective Reinforcement Learning for Online Testing), an online testing technique that combines RL and many-objective search to test the ADS module of an autonomous vehicle.
The authors evaluated MORLOT in the CARLA simulation environment~\cite{Dosovitskiy17}, a widely used high-fidelity driving simulator~\cite{lei-ma-survey}. The ADS under test is the TransFuser model~\cite{Transfuser21}, the highest ranked ADS in the CARLA leaderboard~\cite{carla_leaderboard} at the time of their evaluation.
The results of their empirical evaluation, show that MORLOT outperforms both a random generator, and state-of-the-art search-based techniques in terms of safety requirement violations triggered in the given time budget.

This paper is a replication and extension of the work by Haq \textit{et al}.
Replication of empirical studies is crucial in software engineering research as in ``the construction of knowledge in any empirical science''~\cite{daSilva2014}. As stated by Lindsay and Ehrenberg~\cite{Lindsay93}, replication ``is needed not merely to validate one’s findings, but more importantly, to establish the increasing range of radically different conditions under which the findings hold''.
With this goal, we replicated and then extended the work by Haq \textit{et al}~\cite{Haq23} to investigate the conditions under which RL is actually beneficial in ADS testing. 

The contribution of this paper is twofold:

\begin{description}
    \item [Replication.] The replication does not confirm the finding that many-objective Reinforcement Learning, specifically MORLOT, outperforms the random baseline, in the context of Autonomous Driving System testing. 
    In particular, we reproduce the results of Haq \textit{et al.}~\cite{Haq23}, but show that, if both MORLOT and random are compared in the same conditions, they are statistically indistinguishable. 
    We also comment on the design choices concerning the formulation of the ADS testing problem as an RL problem, and discuss how such choices reduce the learning capability of the RL agent, ultimately making the learning process ineffective with the given time budget.
    \item [Extension.] In the extension study we show that, by formulating the testing problem as single-objective, a deep RL agent converges to an effective policy in most testing scenarios, and significantly outperforms the random baseline.
    The results of our extension study highlight that RL is a promising framework for testing highly dynamic systems such as ADSs, but that further research is needed to address the limitations of the current formulation.
\end{description}

The paper is structured as follows. Section~\ref{sec:background} introduces background and basic definitions. 
Section~\ref{sec:replicated_study} describes the study replicating the work by Haq \textit{et al.}~\cite{Haq23}, while in 
Section~\ref{sec:extension_study} we extend it. 
Section~\ref{sec:related} discusses the related work and highlights the novelty of our contribution. Finally, Section~\ref{sec:conclusions} provides concluding remarks.

%% file: sections/background.tex
\subsection{Reinforcement Learning}
Reinforcement Learning (RL) is the process of learning what to do (i.e., how to relate circumstances to actions) in order to maximize a reward~\cite{Sutton18}. The \textit{learner} (a.k.a., \textit{agent}) is not instructed on actions to take, but needs to explore, interacting with its \textit{environment}, to determine which actions produce the highest reward.

The RL process can be formalized through a Markov Decision Process (MDP), a classical model for sequential decision-making, where actions do not influence just immediate reward, but also the following states. An MDP is defined by a tuple $(S,A,P,R,\gamma)$, where: $S$ and $A$ are respectively the sets of states and actions; $P$ is the state transition probability function $P(s_{t+1}|s_t, a_t)$, assigning the probability of state $s_{t + 1}$ at time step $t + 1$, given state $s_t$ and action $a_t$ at time step $t$; the reward function $R:S \times A \rightarrow \mathbb{R}$ maps a state-action pair to the set of real values; $\gamma \in [0,1]$ is the discount factor controlling the trade-off between future and immediate rewards. 

At time step $t$, the agent observes the state of the environment $s_t$, and selects an action $a_t$ based on its \textit{policy} $\pi$. The policy is generally a stochastic function $\pi: S \rightarrow A$ that yields the probability of selecting action $a_t \in A$ in state $s_t \in S$ at step $t$.
At step $t + 1$ the environment outputs the next state $s_{t+1}$ and a scalar value $r_{t+1}$, rewarding the goodness of action $a_t$. The reward is the learning signal, that the agent aims to maximize. From the continuous interaction with the environment, the agent learns an \textit{optimal policy} $\pi^*$, the one that maximizes the total expected reward the agent gets in its lifetime (the expectation accounts for the randomness of both the transition probability function of the environment and the policy).
The RL methods in the literature differ in how they update the agent's policy as further experience becomes available.

The most common RL algorithms are \textit{model-free} (not equipped with a model of the environment). Within this family, RL algorithms can be categorized into \textit{value-based}, \textit{policy-based}, or a combination of the two. 
Value-based algorithms learn from experience a \textit{value function} giving an estimate of ``how promising'' a state (or a state-action pair) is. The estimate is computed as total expected reward through a \textit{state-value function} $V(s)$ (or an \textit{action-value function} $Q(s,a)$). The policy is then built on the value function, by choosing in each state the action that maximizes it.
Policy-based methods (e.g., policy-gradient) directly maximize the total expected reward by finding a policy through stochastic gradient ascent with respect to the policy parameters (e.g., the parameters of a neural network). 

One of the most important value-based algorithms is \textit{Q-learning}, proposed in 1989 by Watkins~\cite{Watkins89}. It learns an action-value function $Q(s,a)$, updated at every time step as new data becomes available.
The agent's knowledge is represented as a table (named \textit{Q-table}) mapping states and actions to the total expected reward. 
At each time step, the agent starts from state $s_t$, selects the action that has the highest Q-value ($\underset{a \in A}{max} Q(s_t, a)$), transitions into the next state $s_{t+1}$, and collects the reward $r_{t+1}$. Finally, it updates the Q-value of the starting state-action pair $(s_t,a_t)$ with the following formula: \[Q(s_t,a_t) \leftarrow Q(s_t,a_t) + \alpha \left[r_{t+1} + \gamma \underset{a\in A}{max} Q(s_{t+1},a) - Q(s_t,a_t)\right]\]
where: $\gamma$ is the discount factor and $\alpha \in [0,1]$ is the learning rate, controlling the step size at which the Q-values are updated.

Advances in deep learning have led to the development of deep RL (DRL) algorithms like \textit{Deep Q-Network} (DQN)~\cite{mnih2015human}. DQN combines the \textit{Q-learning} paradigm with a neural network that receives a state as input, and approximates the Q-values for each potential action as output. The neural network replaces the Q-table and concisely stores the agent's experience, handling large state spaces such as continuous ones.
To enhance stability during training, DQN typically utilizes a buffer of past experiences. During training, it randomly samples batches of experiences to update the weights of the neural network. 
Additionally, for stabilization DQN uses an auxiliary network (also called target network), a copy of the network being trained. The weights of such network are kept frozen for a certain number of training steps, so that the original network is trained with a fixed target. 
DQN handles continuous state spaces, but still requires actions to be discrete, as its update rule requires a maximization over all the actions for a particular state.

Both Q-learning and DQN use a behavior policy during training to thoroughly explore the environment and look for actions that lead to high reward. A common behavior policy is the $\epsilon$-greedy policy
where the parameter $\epsilon$ represents the probability of choosing a random action instead of relying on the Q-value function (respectively a table and a neural network). Typically, $\epsilon$ is set to 1 at the beginning of training, indicating that the agent always chooses a random action. 
As training progresses and the agent has acquired knowledge of the environment, the $\epsilon$ constant gradually decreases, and the agent starts selecting actions greedily with higher probability. The annealing schedule, i.e., the rate at which the $\epsilon$ constant decreases over time, is problem-specific, as it depends on the rate of exploration that is required to effectively learn a task.

%% file: sections/replicated_study.tex
\subsection{Problem definition}
\label{subsec:problem_def}

In the work by Haq \textit{et al.}~\cite{Haq23}, the ADS is embedded within the CARLA simulator~\cite{Dosovitskiy17}, which renders a realistic town environment including junctions, vehicles, pedestrians, traffic lights, and traffic signs. The ADS that controls the \textbf{ego-vehicle} (\textbf{EV}) has to drive it through a predefined route. At each time step, the ADS receives and processes data from sensors (e.g., camera and LIDAR) to generate driving commands (steering, throttle, and braking) to maximize the driving performance. 
The CARLA leaderboard~\cite{carla_leaderboard} measures the performance of an ADS using the driving score, a combination of two metrics, namely route completion and infraction penalty. The former is defined as the percentage of route completion, while the latter measures the number of infractions during the evaluation, including traffic rules violations (e.g., red lights and stop signs) and detected collisions with other vehicles, pedestrians, and static elements (e.g., road signs).

Haq \textit{et al.} use three test scenarios, one in which the ADS has to drive through a straight road (Straight), one simulating a left turn (Left-Turn), and one for a right turn (Right-Turn). Each scenario has three actors, the EV, a \textbf{vehicle in front} (\textbf{VIF}), and a pedestrian. 

Haq \textit{et al.} define six functional and safety requirements for the ADS: the EV must \textit{(R1)} keep the lane; it must not collide \textit{(R2)} with the VIF, \textit{(R3)} with the pedestrian, and \textit{(R4)} with static meshes (e.g., traffic lights/signs); it must \textit{(R5)} complete the route within the given time, and \textit{(R6)} abide by traffic rules. 

The simulator reports violations of the requirements, respectively: \textit{(V1)} if the \textbf{distance from the center of the lane} (\textbf{DCL}) exceeds a threshold identifying the lane boundaries; \textit{(V2)} if the \textbf{distance from the VIF} (\textbf{DV}), or \textit{(V3)} the \textbf{distance from the pedestrian} \textbf{(DP)}, or \textit{(V4)} \textbf{distance from static meshes} \textbf{(DS)} is less than or equal to zero (i.e., a collision occurs); 
\textit{(V5)} if at the end of the scenario the \textbf{distance from the destination} \textbf{(DT)} is greater than zero,
and \textit{(V6)} if it detects that the EV has violated a \textbf{traffic rule} \textbf{(TR)} (e.g., running a red light). MORLOT~\cite{Haq23} changes the environment in each route to find violations of the requirements.

The authors formulate testing as an RL problem. %
The state space has 19 variables specifying the position, speed, and acceleration of the EV and the VIF, the position and speed of the pedestrian, and environmental conditions (i.e., fog and rain intensity, and sun altitude). The action space of the RL agent is discrete: the agent can choose in a set of 17 actions controlling the VIF (throttle and steering), the pedestrian (speed and position), and the environmental conditions. 
Each action changes the value of a controlled variable by a small and constant amount (e.g., the action indices for the throttle of the VIF increase/decrease by 0.1 the current throttle value). 
Finally, a reward function is defined for each requirement. The reward function for the R6/TR requirement is binary: the agent gets a large reward when the requirement is violated, zero otherwise. For all other requirements, the reward is a function of the \texttt{distance} to the violation. Specifically, the reward is equal to $\frac{1}{\texttt{distance}}$ when $\texttt{distance} > 0$, and to $1.0$E$+06$ otherwise (a requirement is violated). The testing goal is to minimize the distances DV, DP, and DS, and maximize the distances DCL and DT (in such cases the authors subtract 1 from the original distance value, i.e., $1 - \texttt{distance}$).\footnote{DCL and DT range between 0 and 1.}

MORLOT is implemented as a many-objective search with Q-learning. In particular, MORLOT builds a Q-table for each requirement to be violated, such that each Q-table is updated with the transitions (i.e., $\langle$\textit{state, action, reward, next state}$\rangle$ tuples) related to the respective reward function. In this way, each Q-table learns the optimal state-action pairs to violate the respective requirement. MORLOT stores a list of uncovered objectives (initially all of them are uncovered), and at each step selects the action from the Q-table associated with the uncovered objective closest to being covered (i.e., the objective with the highest reward at the previous step). 

\subsection{Subject and configuration}

The ADS under test is TransFuser~\cite{Transfuser21}, a Deep Neural Network model submitted to the CARLA Autonomous Driving challenge. TransFuser is a Multi-Modal Fusion Transformer to predict the trajectory of the ego vehicle to determine the driving commands.

Haq \textit{et al.} used the first version of TransFuser, proposed in 2021~\cite{Transfuser21}. The code of the TransFuser agent, and the pre-trained models, are open-source.\footnote{\url{https://github.com/autonomousvision/transfuser/tree/cvpr2021}.} The replication package of MORLOT\footnote{Figshare DOI: 10.6084/m9.figshare.20526867.v2.} links to the same models and contains the three scenarios that the authors used for their evaluation (Straight, Left-Turn, Right-Turn).

We configured the environment to run TransFuser following the instructions in the replication package of MORLOT, as well as the instructions by the authors of TransFuser.
We validated our local configuration of the ADS by running the agent in the three scenarios and checking that no violation occurred (i.e., the TransFuser agent drives well in nominal conditions). 

\subsection{Replication}
\subsubsection{Methodology}

The replication package of Haq \textit{et al.} provides the code to execute the considered test generators, after configuring the ADS and the simulator with the given scenario. By inspecting the code in the package, we found an important, initially undocumented, configuration option 
(later added by the authors in the package README file), 
a boolean flag named "RL", that determines the way violations are detected.
As the name suggests, it needs to be set to true when running RL-based test generators (like MORLOT), while it needs to be set to false when running the random baseline, to allow the use of simulator sensors for detecting collisions and lane invasions. When running the random baseline with the flag set to true, the detection is solely based on distances between objects given by the simulator. We execute our experiments with both versions of random. We call RANDOM\_False the baseline with the flag set to false, and RANDOM\_True, the baseline with the flag set to true (i.e., the default value in the original replication package).

We compare MORLOT to the baselines in terms of effectiveness (coverage of requirements) and efficiency (coverage achieved over time), following the authors' evaluation~\cite{Haq23}. The coverage of a technique is 100\% when it finds at least one violation per requirement within the time budget of 4 hours, as in the original study. MORLOT adopts an $\epsilon$-greedy strategy with $\epsilon$ linearly decreasing from 1 to 0.1 in the first 48 min\-utes (20\% of the budget).
To account for randomness, we executed each technique 20 times (twice the repetitions used by Haq \textit{et al.}) in each of the three scenarios, for a total of 720 hours of computation. All experiments were executed on the Google Cloud Compute Engine platform %
on a virtual machine with Ubuntu 18.04, an Intel Haswell CPU (4 cores) and an NVIDIA Tesla T4 with 16 GB of VRAM.

\subsubsection{Effectiveness}

\begin{figure}[t]
	\centering
	\includegraphics[width=0.94\columnwidth]{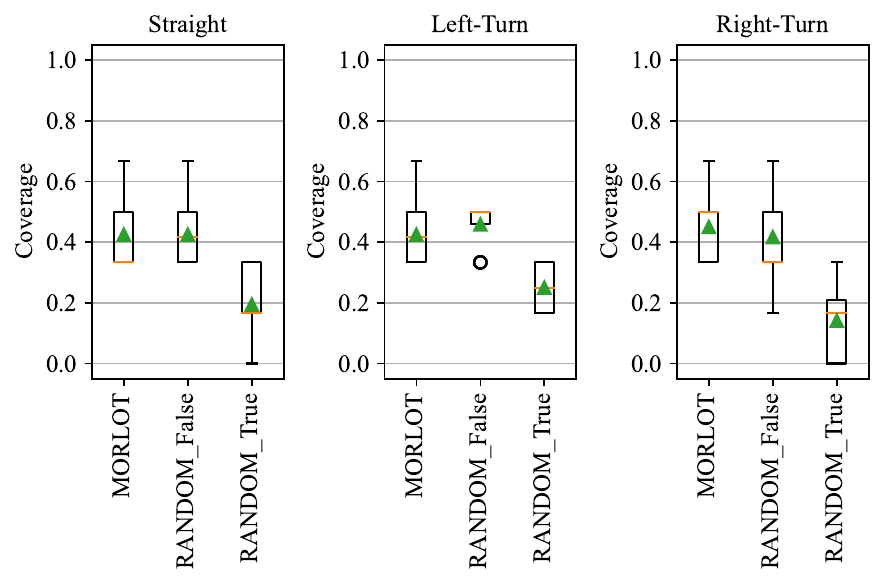}
    \caption{Coverage of safety and functional requirements} 
	\label{fig:rep_effectiveness}
\end{figure}

\begin{table}[b]
\centering
\caption{Pairwise comparisons Dunn Test - Coverage of the requirements (statistically significant differences are in bold).} 
\label{tab:rep_dunn}
\resizebox{\columnwidth}{!}{%
\renewcommand{\arraystretch}{1.02}
\setlength{\dashlinedash}{0.8pt}\setlength{\dashlinegap}{2.4pt}
\begin{tabular}{c|ccc}
Techniques & Straight & Left-Turn & Right-Turn \\ \hline
$\bigtriangleup$MORLOT vs. RANDOM\_False$\bigtriangledown$ & 1.00E+00 & 8.61E-01 & 1.00E+00 \\ \hdashline
$\bigtriangleup$MORLOT vs. RANDOM\_True$\bigtriangledown$ & $\bigtriangleup$\textbf{\textless{}1.00E-04} & $\bigtriangleup$\textbf{\textless{}1.00E-04} & $\bigtriangleup$\textbf{\textless{}1.00E-04} \\ \hdashline
$\bigtriangleup$RANDOM\_False vs. RANDOM\_True$\bigtriangledown$ & $\bigtriangleup$\textbf{\textless{}1.00E-04} & $\bigtriangleup$\textbf{\textless{}1.00E-04} & $\bigtriangleup$\textbf{\textless{}1.00E-04} \\ \bottomrule
\end{tabular}}
\end{table}

Figure~\ref{fig:rep_effectiveness} shows the coverage achieved by the three testing techniques over the 20 repetitions. The orange line and the green arrow represent the median and the mean coverage, respectively. 
Results show that RANDOM\_True achieves the lowest coverage as it never covers more than 2 requirements (i.e., 0.33). On the other hand, MORLOT and RANDOM\_False exhibit a similar average coverage. MORLOT's median is slightly lower than RANDOM\_False's in Straight and Left-Turn routes but higher in the Right-Turn route. Both techniques consistently cover at least 2 requirements, with a maximum of 4 (i.e., 0.66).

We run the Friedman test~\cite{Friedman1937}, a non-parametric hypothesis test for Analysis of Variance, to assess if there is at least one testing technique that significantly differs from the others. The Friedman test detects a significant difference for at least one pair for each route (the \textit{p}-values are, respectively, 3.93E-06, 1.734E-06, 1.458E-08, for Straight, Left-Turn, and Right-Turn). 

We run the Dunn test~\cite{Dunn64} for \textit{post hoc} analysis to detect which pair of techniques differ significantly. Table~\ref{tab:rep_dunn} reports the \textit{p}-values for each pair. The test confirms that there is no significant difference between MORLOT and RANDOM\_False, while both of them show a statistically higher coverage than RANDOM\_True. 

We also compared the testing techniques based on the average number of violations per requirement they trigger. Table~\ref{tab:rep_avg_violations} shows that MORLOT and RANDOM\_False trigger a comparable number of violations (respectively: 6.15 and 6.2 for Straight, 16.35 and 15.35 for Left-Turn, 5.45 and 5.3 for Right-Turn), with both being better than RANDOM\_True in all the three routes (it only triggers 1.7, 5.5, and 1.1 violations for Straight, Left-Turn, and Right-Turn respectively).

Both the coverage of the safety requirements and the average number of violations show that activating sensors to detect collisions and lane invasions (i.e., setting the ``RL'' boolean configuration flag to true) increases the measured effectiveness of the RANDOM\_False baseline. Indeed, we noticed, by inspecting the execution logs, that most of the times a collision is detected by sensors, the distance value is close to the threshold but does not exceed it (the threshold value is 0 in the case of the DV, DP, DS requirements and equal to 1.15 in the case of DCL). This issue affects the RANDOM\_True baseline, significantly decreasing its measured effectiveness, but does not represent an issue for the reward computation for MORLOT, since the implementation forces the distance value to the threshold value whenever a sensor detects a violation.

\begin{table*}[]
\centering
\caption{Average number of violations found by MORLOT, RANDOM\_False (RAND\_F), and RANDOM\_True (RAND\_T).} 
\label{tab:rep_avg_violations}
\setlength{\tabcolsep}{1.6pt}
\setlength{\dashlinedash}{0.8pt}\setlength{\dashlinegap}{2.4pt}
\resizebox{\columnwidth}{!}{%
\begin{tabular}{l|ccc|ccc|ccc}
\hline
 & \multicolumn{3}{c|}{Straight} & \multicolumn{3}{c|}{Left-Turn} & \multicolumn{3}{c}{Right-Turn} \\ \cline{2-10} 
Requirement & MORLOT & RAND\_F & RAND\_T & MORLOT & RAND\_F & RAND\_T & MORLOT & RAND\_F & RAND\_T \\ \hline
R1, DCL & 0.15 & 0.20 & 0.15 & 5.85 & 4.90 & 4.90 & 0.00 & 0.00 & 0.00 \\ \hdashline
R2, DV & 1.40 & 1.50 & 0.00 & 9.60 & 9.20 & 0.00 & 0.60 & 0.45 & 0.00 \\ \hdashline
R3, DP & 3.80 & 3.85 & 0.80 & 0.85 & 1.25 & 0.60 & 3.05 & 3.10 & 0.45 \\ \hdashline
R4, DS & 0.00 & 0.00 & 0.00 & 0.00 & 0.00 & 0.00 & 0.00 & 0.00 & 0.00 \\ \hdashline
R5, DT & 0.75 & 0.65 & 0.75 & 0.00 & 0.00 & 0.00 & 0.55 & 0.55 & 0.65 \\ \hdashline
R6, TR & 0.05 & 0.00 & 0.00 & 0.05 & 0.00 & 0.00 & 1.25 & 1.20 & 0.00 \\\hline
TOTAL & 6.15 & 6.20 & 1.70 & 16.35 & 15.35 & 5.50 & 5.45 & 5.30 & 1.10 \\\hline
\end{tabular}}
\end{table*}

\subsubsection{Efficiency}

\begin{figure*}[t]
	\centering
	\includegraphics[width=0.99\textwidth]{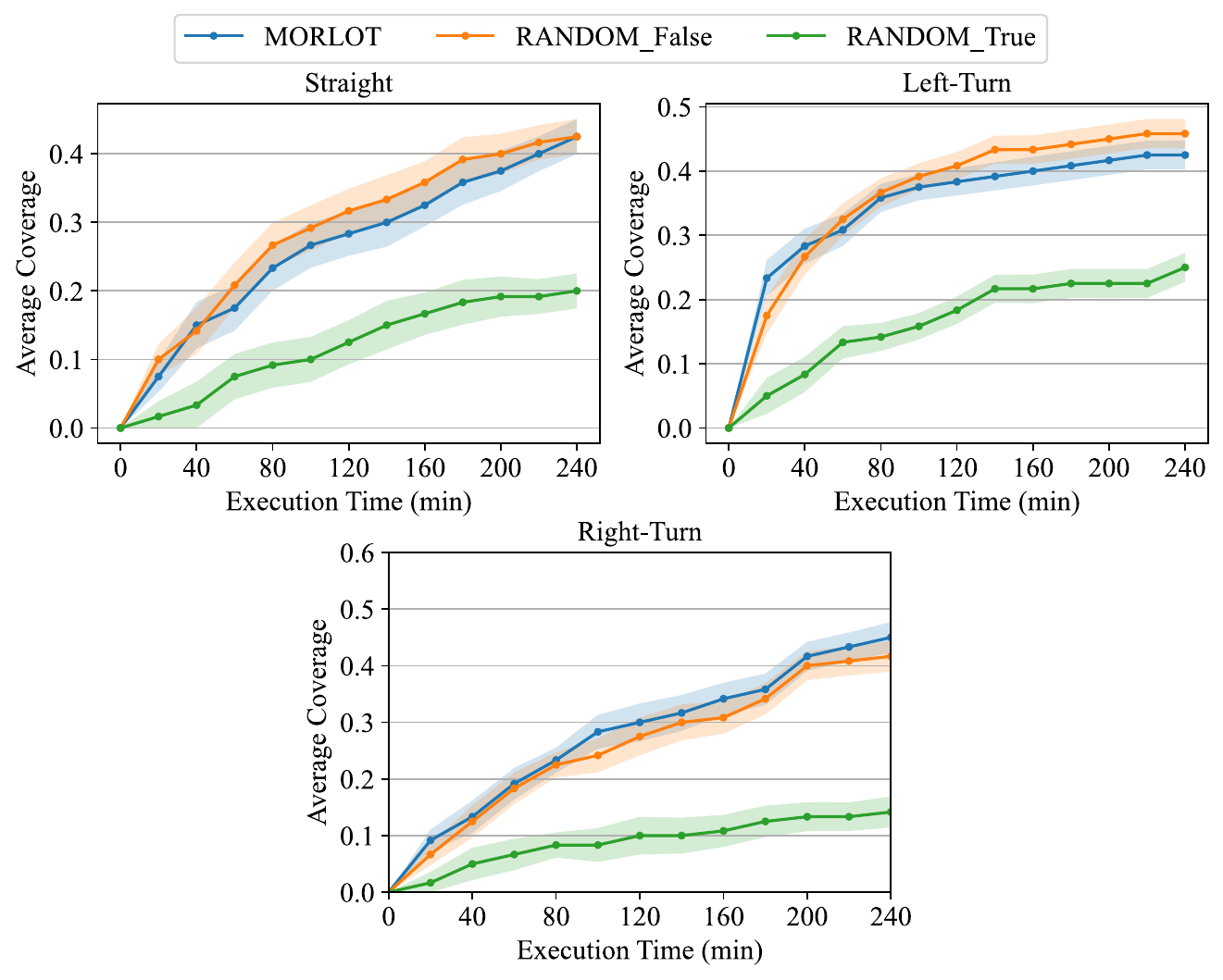}
\caption{Average coverage of safety and functional requirements over time.} 
	\label{fig:rep_efficiency}
\end{figure*}

We compare the testing techniques also for their efficiency, i.e., the coverage of the requirements over time.
Figure~\ref{fig:rep_efficiency} shows the trend of average coverage per technique in the 20 repetitions. Each average coverage value is sampled every 20 minutes; the shaded area represents the standard error of the mean (i.e., $\pm \, s/\sqrt{n}$).
RANDOM\_True always achieves the lowest coverage on all routes. In the Straight and Left-Turn routes, it achieves only half the coverage of MORLOT and RANDOM\_False, and only at the end of the 4-hour budget; in the Right-Turn route it achieves only a third of the coverage. On the other hand, RANDOM\_False and MORLOT achieve comparable coverage over time. MORLOT curve mostly stays slightly below the coverage curve of the RANDOM\_False baseline, with the exception of the Right-Turn route. 

We computed the Area Under the Curve (AUC), for the curves that plot the coverage of the techniques over time (Figure~\ref{fig:rep_efficiency}). Table~\ref{tab:rep_avg_area} reports the (min-max)\footnote{With min and max respectively equal to 0 and the overall maximum value.} normalized mean value and standard deviation for all routes. Again, RANDOM\_True achieves significantly lower values, while MORLOT and RANDOM\_False obtain comparable results, with the first being slightly worse among all routes except for Right-Turn. The Friedman test applied to the AUC values detects a significant difference for at least one pair for each route (the \textit{p}-values are respectively 1.09E-05, 4.11E-10, 4.11E-10 for Straight, Left-Turn, and Right-Turn). The Dunn Test (Table~\ref{tab:rep_dunn_area}) confirms that MORLOT and RANDOM\_False are equivalent, while both outperform RANDOM\_True.

\begin{table}[]
\addtocounter{table}{1}
\centering
\caption{Area under the curve: Normalized Mean and Std.} 
\label{tab:rep_avg_area}
\renewcommand{\arraystretch}{1.02}
\setlength{\dashlinedash}{0.8pt}\setlength{\dashlinegap}{2.4pt}
\begin{tabular}{c|ccc}
 & Straight & Left-Turn & Right-Turn \\ 
 Technique & Mean ($\pm$Std) & Mean ($\pm$Std) & Mean ($\pm$Std) \\ \hline
MORLOT  & 0.52 ($\pm$0.22) & 0.64 ($\pm$0.12) & 0.60 ($\pm$0.18) \\ \hdashline
RANDOM\_False & 0.57 ($\pm$0.20) & 0.67 ($\pm$0.12) & 0.56 ($\pm$0.21) \\ \hdashline
RANDOM\_True & 0.23 ($\pm$0.18) & 0.30 ($\pm$0.09) & 0.19 ($\pm$0.20) \\ \bottomrule
\end{tabular}
\end{table}

\begin{table}[]
\centering
\caption{Area under the curve: Pairwise comparisons Dunn Test (statistically significant differences are in bold).} 
\label{tab:rep_dunn_area}
\renewcommand{\arraystretch}{1.1}
\resizebox{\columnwidth}{!}{%
\setlength{\dashlinedash}{0.8pt}\setlength{\dashlinegap}{2.4pt}
\begin{tabular}{c|ccc}
\setlength{\tabcolsep}{1.55pt}
Techniques & Straight & Left-Turn & Right-Turn \\ \hline
$\bigtriangleup$MORLOT vs. RANDOM\_False$\bigtriangledown$ & 1.00E+00 & 1.00E+00 & 1.00E+00 \\ \hdashline
$\bigtriangleup$MORLOT vs. RANDOM\_True$\bigtriangledown$ & $\bigtriangleup$\textbf{1.00E-03} & $\bigtriangleup$\textbf{\textless{}1.00E-04} & $\bigtriangleup$\textbf{\textless{}1.00E-04} \\ \hdashline
$\bigtriangleup$RANDOM\_False vs. RANDOM\_True$\bigtriangledown$ & $\bigtriangleup$\textbf{\textless{}1.00E-04} & $\bigtriangleup$\textbf{\textless{}1.00E-04} & $\bigtriangleup$\textbf{1.00E-04} \\ \bottomrule
\end{tabular}}
\end{table}

\subsection{Discussion}

By comparing our findings with those reported by Haq \textit{et al.}, we observe that MORLOT and RANDOM\_True match the results of the original paper. However, upon setting the sensor activation flag to false in the random baseline (which we call RANDOM\_False in the evaluation), we found no significant difference with MORLOT, both in terms of effectiveness and efficiency. We investigated the causes for the poor performance of MORLOT w.r.t. RANDOM\_False and we identified two main reasons, related to the design and implementation of the RL algorithm.

\begin{figure*}[]
	\centering
	\includegraphics[width=0.99\textwidth]{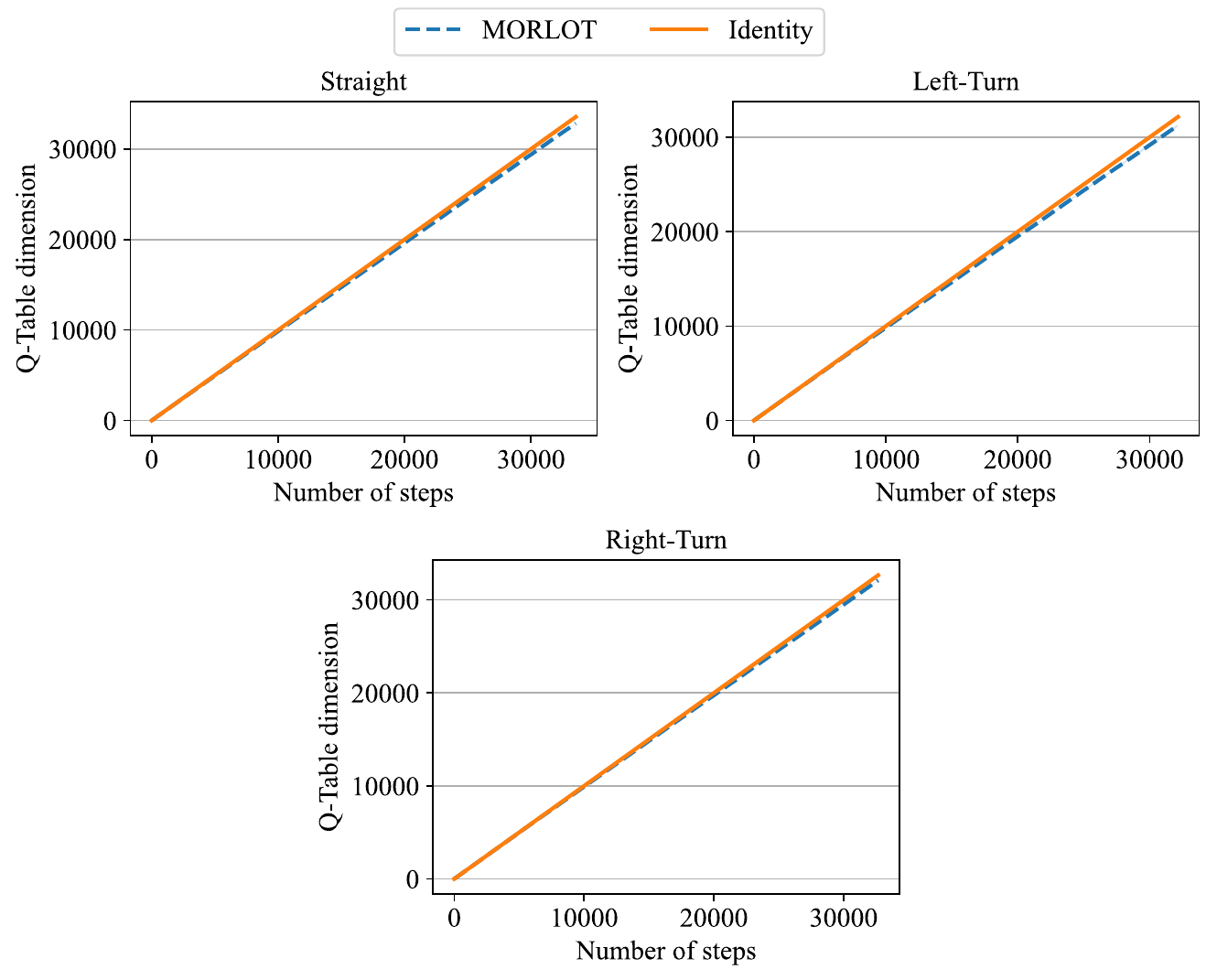}
\caption{Average Q-Table dimension (number of observed states) over number of steps.} 
	\label{fig:rep_qt_dim}
\end{figure*}

The first reason regards Q-learning, the RL algorithm chosen by Haq \textit{et al.}. Q-learning requires both state and action spaces to be discrete.
In MORLOT, actions are discretized, but the state is implemented as a string concatenating the continuous values of the 19 state variables. 
MORLOT's implementation of Q-learning is dynamic, i.e., it initializes an empty table and adds a row each time it encounters a new state. During training, if it revisits a previously discovered state, it updates the corresponding Q-value. To explore the evolution of the Q-table dimension, we monitored the count of distinct states memorized in the Q-table (i.e., its dimension) relative to the total number of steps executed by the agent in a single repetition. To prevent potential biases due to the additional computation given by the collection this information, we ran 3 additional repetitions for MORLOT in the three routes. The average Q-table dimensions are depicted in Figure \ref{fig:rep_qt_dim}, revealing a nearly linear progression over time.
This suggests that the agent rarely encounters the same state more than once, decreasing the learning effectiveness of the algorithm and its ability to detect violations.

The second reason concerns MORLOT's definition of the reward functions and the way the algorithm uses the Q-table to choose the action at each time step. 
As for requirement R6 on traffic rules (TR), the reward function is sparse (reward is different from $0$ only in case of vio\-lations, which are very rare).
For requirement R5 on distance from destination DT, the reward function starts at its maximum at the beginning of the route (maximum DT) and decreases as the EV progresses along the route. The function rewards the RL agent, independently\footnote{An exception is when the VIF stops and the EV waits for it to move out of its trajectory. In this case, the reward remains constant.} of the chosen action, which makes it difficult for the RL algorithm to properly assign credit to the actions that make the ADS violate R5.
As for requirements R1--R4 (on DCL, DV, DP, DS, respectively), some of them are trivial to violate, while others are very challenging (e.g., R4/DS is rarely violated). We noticed that MORLOT covers first the easiest requirements to violate (namely: R2/DV and R3/DP in Straight; R1/DCL and R2/DV in Left-Turn)\footnote{We also observed that most of the times R2/DV is violated, R1/DCL is violated, too.} and then starts to select actions from the Q-table associated with the R4/DS requirements. 
This happens because DS has one of the highest average reward values, and MORLOT selects the Q-table with the highest reward at each step. Since R4/DS is difficult to violate, this results in MORLOT wasting the remaining search budget without addressing any additional requirement. 
Table~\ref{tab:rep_rew_val} shows the mean and standard deviations obtained when running MORLOT, for each reward function associated with the respective requirement.\footnote{TR is not reported as it is binary.} Notably, DS ranks as the third-highest in terms of reward value.

\begin{table}[!ht]
\centering
\caption{Statistics of rewards per requirement.} 
\label{tab:rep_rew_val}
\renewcommand{\arraystretch}{1.08}
\setlength{\dashlinedash}{0.8pt}\setlength{\dashlinegap}{2.4pt}
\begin{tabular}{c|ccccc}
 & R1/DCL & R2/DV & R3/DP & R4/DS & R5/DT \\ \hline
Mean (Median) & 1.7 (1.1) & 12.1 (11.1) &9.0 (5.9) & 6.8 (5.3) & 0.8 (0.9) \\ \hdashline
Std (IQR) & 37.4 (0.2) & 7.4 (11.1) & 9.5 (6.0) & 4.3 (5.64) & 0.2 (0.33) \\ \bottomrule
\end{tabular}
\end{table}

%% file: sections/extension_study.tex
The results of the replication study motivate to further investigate the extent to which RL can benefit online testing of ADS.
The extension uses DQN and considers the reward function for only one of the requirements, DV. Indeed, DQN can naturally handle continuous state spaces and is expected to scale to the size of the case study's state space. The reason for choosing DV for the refinement of the reward function 
is that it proved to be a non-trivial yet coverable requirement in all routes. 

We compare Q-learning and DQN against the RANDOM\_False baseline (called simply RANDOM hereinafter) on the same tasks defined by Haq \textit{et al.}
As for the replication study, we execute 20 repetitions with 4 hours of budget for all techniques on the three routes (Straight, Left-Turn, and Right-Turn), for 720 hours of computation.

\subsection{Problem definition}
The replicated study showed that: (1) inclusion of multiple requirements, some of which are almost impossible to violate, but still return high rewards, is one of the reasons for the degenerate behavior of RL, which performed comparably to random; (2) adoption of Q-learning reduces the agent's learning capability, as similar continuous states are not recognized as recurring states. Hence, in our extension study, we focus on a single requirement, i.e., \textit{respect safety distance from other vehicles}, and compare two RL methods (i.e., DQN and Q-learning) with RANDOM as baseline.\footnote{The hyperparameters used for DQN are reported in our replication package: \url{https://doi.org/10.6084/m9.figshare.24794544}.} 
Existing studies show that this requirement covers the majority of challenging situations for an ADS, as more than 80\% of the accidents involving an ADS in California are caused by the maneuvers of other vehicles~\cite{favaro-crashes-california,nhtsa-summary-crashes}.

We also update the termination condition of an episode, to take into account the achievement of the driving objective. An episode stops both when the simulator detects a collision between the ADS and the vehicle in front (a violation is detected), and when the ADS overtakes the vehicle in front (a violation can no longer occur).

The \textit{goal} of the study is to assess if an RL agent trained with DQN %
performs better 
than RANDOM under the settings of Haq \textit{et al.} (see Section~\ref{subsec:problem_def}), in a single-objective problem formulation.

\subsection{Experiments}
\subsubsection{Evaluation criteria}
We compare DQN, Q-learning (simply Q, hereinafter), and RANDOM in terms of \textit{effectiveness} and \textit{efficiency}. %
Effectiveness is computed as the total number of violations of the R2/DV requirement (collisions with VIF) in the time budget of 4 hours. 
Efficiency is computed as the average number of violations found over time. Then, in a \textit{qualitative evaluation}, we analyze the failing scenarios detected by the competing techniques.

\subsubsection{Effectiveness}
\begin{figure}[t]
	\centering
	\includegraphics[width=0.96\columnwidth]{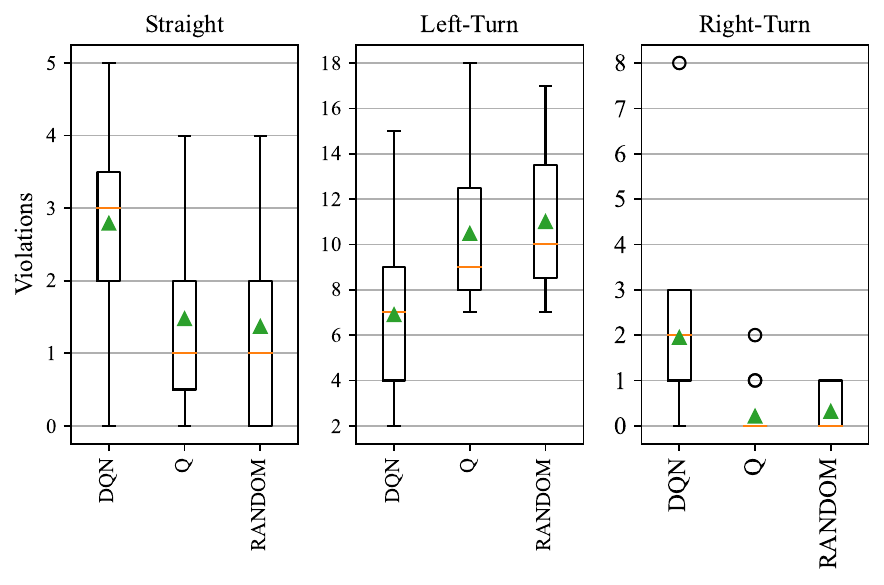}
	\caption{Box plot of the distribution of the number of violations of requirement R2 (collisions with Vehicle in Front).} 
	\label{fig:ext_effectiveness}
\end{figure}

Figure~\ref{fig:ext_effectiveness} shows a box plot of the number of violations of the techniques in the 20 repetitions. We observe that Q triggers as many violations as the RANDOM baseline across all routes. This supports the conclusions of the replication study, indicating that Q, even with a single-objective formulation, is indistinguishable from RANDOM.
On the contrary, a deep RL agent like DQN demonstrates superior performance compared to the other two techniques in most of the routes (i.e., Straight and Right-Turn), where it triggers, on average, twice the number of violations compared to Q and RANDOM. On the other hand, DQN is less effective than the competing techniques in the Left-Turn route.

\begin{table}[b]
\centering
\caption{Pairwise comparisons Dunn Test - Number of violations (statistically significant differences are in bold).} 
\label{tab:ext_dunn}
\renewcommand{\arraystretch}{1.02}
\setlength{\dashlinedash}{0.8pt}\setlength{\dashlinegap}{2.4pt}
\begin{tabular}{c|ccc}
\setlength{\tabcolsep}{1.pt}
Techniques & Straight & Left-Turn & Right-Turn \\ \hline
$\bigtriangleup$DQN vs. Q$\bigtriangledown$ & $\bigtriangleup$\textbf{3.43E-02} & $\bigtriangledown$\textbf{9.20E-03} & $\bigtriangleup$\textbf{\textless{}1.00E-04} \\ \hdashline
$\bigtriangleup$DQN vs. RANDOM$\bigtriangledown$ & $\bigtriangleup$\textbf{4.00E-03} & $\bigtriangledown$\textbf{1.80E-03} & $\bigtriangleup$\textbf{2.90E-03} \\ \hdashline
$\bigtriangleup$Q vs. RANDOM$\bigtriangledown$ & 1.00E+00 & 1.00E+00 & 1.00E+00 \\ \bottomrule
\end{tabular}
\end{table}

\begin{figure*}[b]
	\centering
	\includegraphics[width=0.94\textwidth]{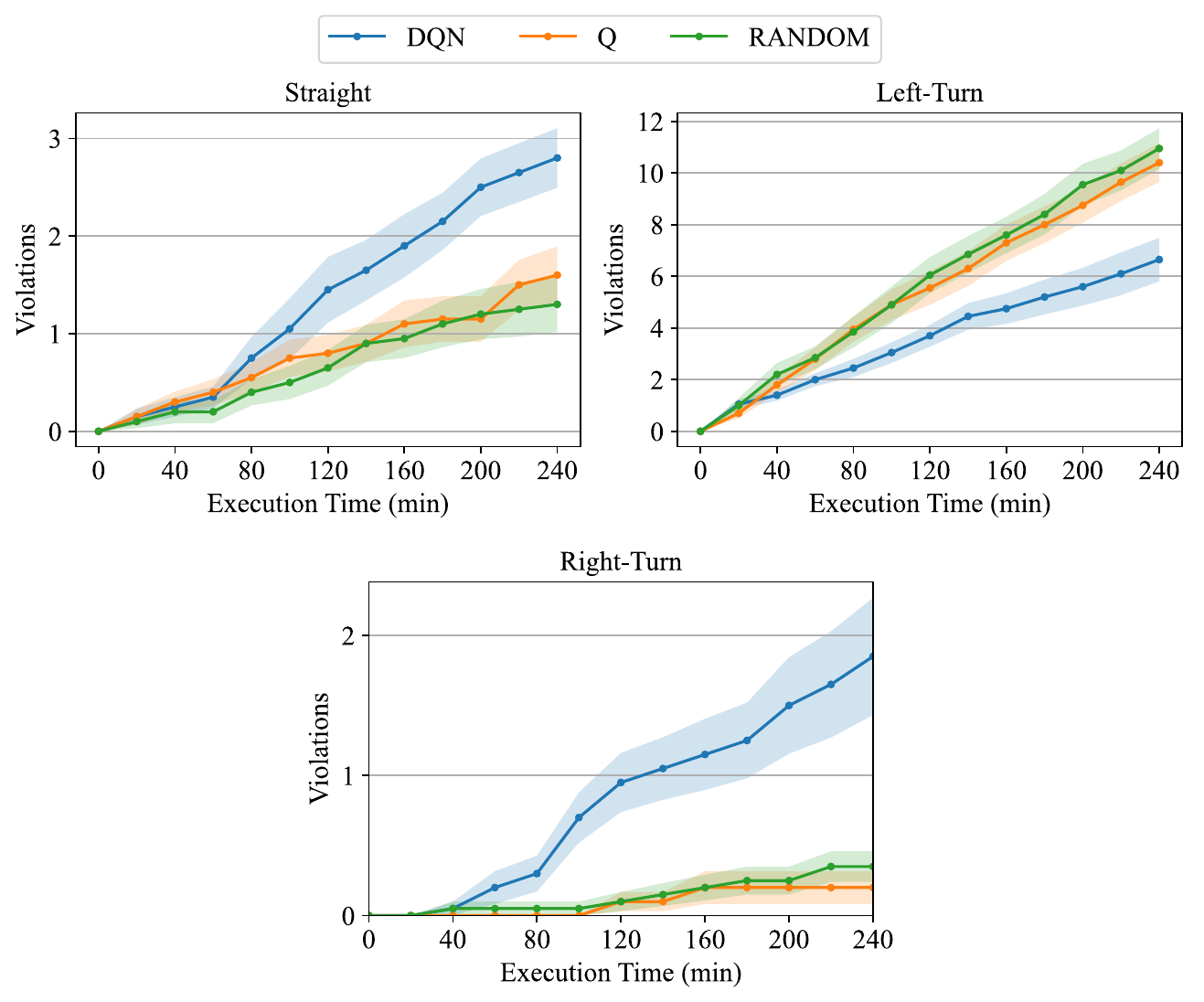}
\caption{Average number of violations (i.e., collision with vehicles) over time.} 
	\label{fig:ext_efficiency}
\end{figure*}

To statistically compare the three techniques, we run the Friedman test, which detects a significant difference for at least one pair of approaches for all the routes (\textit{p}-value = 1.94E-03, 7.76E-03, 1.67E-03). Table~\ref{tab:ext_dunn} reports the \textit{p}-values for all the pairwise comparisons, computed with the Dunn test. DQN significantly outperforms both Q and RANDOM in the Straight and Right-Turn routes, while performing worse than both in the Left-Turn route. On the other hand, Q and RANDOM are statistically indistinguishable.

\subsubsection{Efficiency}
Figure~\ref{fig:ext_efficiency} shows the average number of violations identified by the three techniques over time, with the shaded area representing the standard error of the mean (i.e., $\pm \, s/\sqrt{n}$). We observe that DQN is the most efficient technique in 2 of the 3 routes. For the Straight route, DQN triggers the first violation, on average, one hour earlier than  Q and RANDOM, and it requires less than half the budget to obtain the same number of violations. In the Right-Turn route, DQN stands out as the only technique consistently able to discover up to two violations in each repetition. 
In the Left-Turn route, DQN is less efficient than Q and RANDOM. 

Q and DQN have the same trend as RANDOM in the initial 50 minutes, that is the time budget allocated for $\epsilon$-greedy exploration. 
However, unlike Q, which maintains the same behavior as RANDOM through the entire time budget, DQN is able to learn an effective policy to trigger violations of the ADS under test. 

To quantitatively evaluate the efficiency, we measured the AUC of the average number of violations over time (Figure~\ref{fig:ext_efficiency}). Table~\ref{tab:ext_avg_area} shows the (min-max) normalized AUC mean and standard deviation per technique, in all routes. DQN covers 54\% of the area in Straight and 48\% in Right-Turn, while the coverage is lower than Q and RANDOM in Left-Turn (i.e., 36\% vs $\approx$ 56\% of Q and RANDOM). The Friedman test applied to the AUC values detects a significant difference for at least one pair for Left-Turn and Right-Turn routes, but not for the Straight route (i.e., the \textit{p}-values are 1.10E-01, 1.82E-02, and 5.88E-03 for Straight, Left-Turn, and Right-Turn, respectively).

\begin{table}[t]
\centering
\caption{Area under the curve: Normalized Mean and Std.} 
\label{tab:ext_avg_area}
\renewcommand{\arraystretch}{1.02}
\setlength{\dashlinedash}{0.8pt}\setlength{\dashlinegap}{2.4pt}
\begin{tabular}{c|ccc}
Technique & Straight & Left-Turn & Right-Turn \\ \hline
DQN  & 0.38 ($\pm$0.26) & 0.32 ($\pm$0.15) & 0.34 ($\pm$0.32) \\ \hdashline
Q & 0.22 ($\pm$0.20) & 0.48 ($\pm$0.20) & 0.03 ($\pm$0.09) \\ \hdashline
RANDOM & 0.19 ($\pm$0.19) & 0.51 ($\pm$0.22) & 0.06 ($\pm$0.10) \\ \bottomrule
\end{tabular}
\end{table}

We then ran the Dunn test for all pairs of approaches and for the routes in which Friedman exposed a significant difference (i.e., Left-Turn and Right-Turn).
Table~\ref{tab:ext_dunn_area} shows that there is no significant difference between Q and RANDOM in all routes. DQN shows a higher AUC than both in the Right-Turn route. 
Despite no statistically significant difference in the Straight route, the trend is in favor of DQN (on average, the AUC values of DQN compared to Q and RANDOM are respectively 53\% vs 31\% and 26\%). In the Left-Turn route, DQN has an AUC lower than both Q and RANDOM.

\begin{table}[t]
\centering
\caption{Area under the curve - Pairwise comparison Dunn Test (statistically significant differences are in bold).} 
\label{tab:ext_dunn_area}
\setlength{\dashlinedash}{0.8pt}\setlength{\dashlinegap}{2.4pt}
\begin{tabular}{c|ccc}
Techniques & Straight & Left-Turn & Right-Turn \\ \hline
$\bigtriangleup$DQN vs. Q$\bigtriangledown$ & - & $\bigtriangleup$\textbf{2.30E-02} & $\bigtriangleup$\textbf{1.00E-04} \\ \hdashline
$\bigtriangleup$DQN vs. RANDOM$\bigtriangledown$ & - & $\bigtriangleup$\textbf{1.13E-02} & $\bigtriangleup$\textbf{3.50E-03} \\ \hdashline
$\bigtriangleup$Q vs. RANDOM$\bigtriangledown$ & - & 1.00E+00 & 1.00E+00 \\ \bottomrule
\end{tabular}
\end{table}

\begin{figure*}[t]
	\centering
	\includegraphics[width=0.94\textwidth]{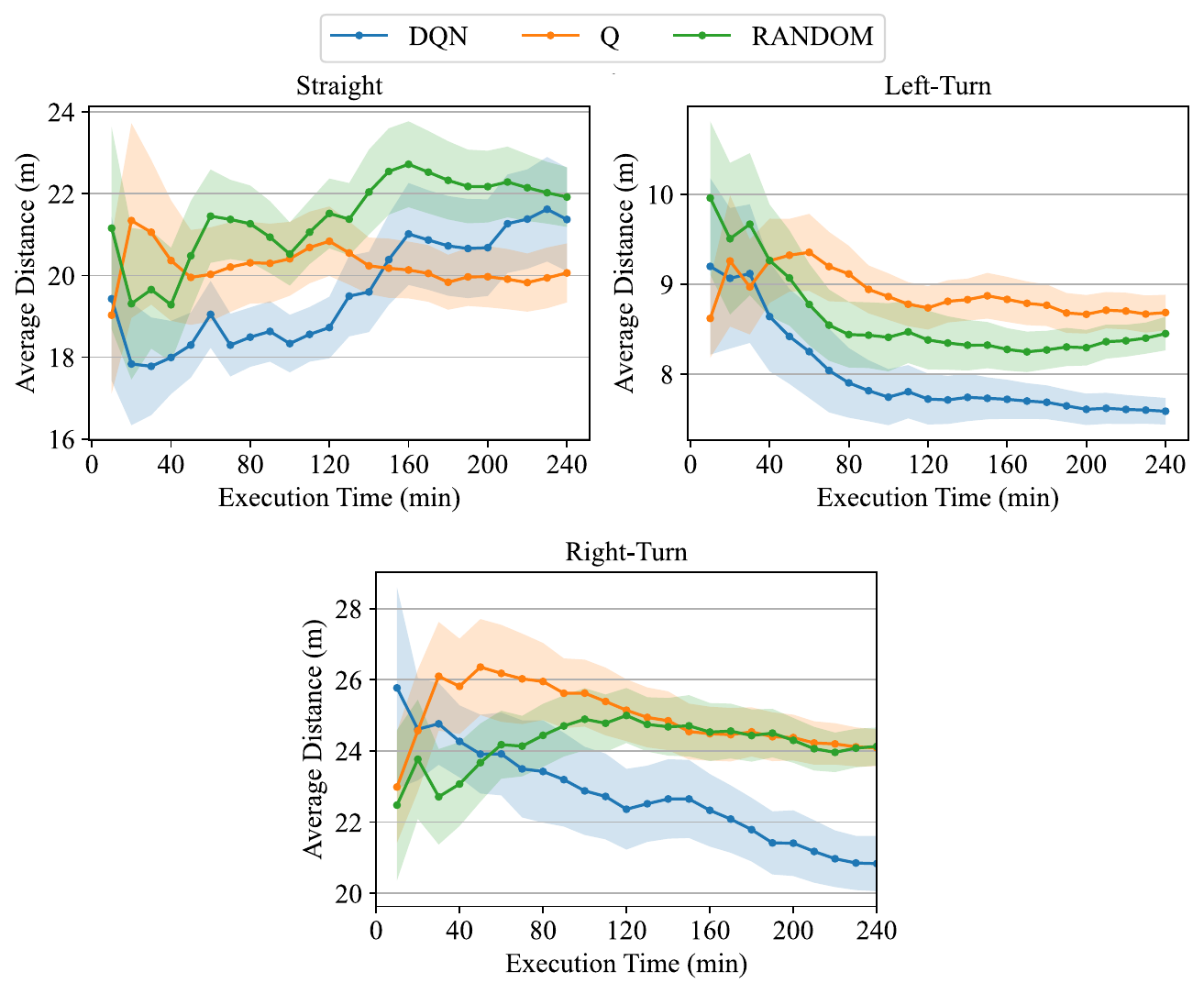}
    \caption{Average distance from the Vehicle in Front (DV) over time (10 minutes window).} 
	\label{fig:ext_avg_distance}
\end{figure*}

We investigate the performance difference of DQN in the three scenarios, by first analyzing the trend of the metric used to quantify the distance from the vehicle in front (DV) requirement over time. Since the reward function for Q and DQN is the inverse of the DV metric (until a requirement violation occurs, when a reward of $1.0$E$+06$ is returned), the RL agent aims at minimizing the distance between the ego vehicle (EV) and the vehicle in front (VIF). 
Figure~\ref{fig:ext_avg_distance} shows the trend of this distance (in meters), where each point represents an average of the distance values collected in a window of 10 minutes of execution; we then average all the values across the 20 repetitions.

In two scenarios - Left-Turn and Right-Turn - DQN makes the distance decrease over time (i.e., it maximizes the average reward); however, comparing the average distance over time (Figure~\ref{fig:ext_avg_distance}) and the average number of violations over time (Figure~\ref{fig:ext_efficiency}), we notice that in one of the two scenarios (Left-Turn) a decreasing distance does not correspond to a higher number of violations. Moreover, in the Straight route, DQN outperforms both Q and RANDOM for number of violations, while the three seem to be equivalent w.r.t. the distance trend (left-most plots in Figures~\ref{fig:ext_efficiency}, \ref{fig:ext_avg_distance}). 
This might be due to the effectiveness of the large reward returned upon collision.
In the Left-Turn route, DQN finds less violations than Q and RANDOM despite being able to minimize the distance between the two vehicles over time (center plots in Figures~\ref{fig:ext_efficiency}, \ref{fig:ext_avg_distance}). The only scenario where the number of violations triggered by DQN increases as the distance decreases is the Right-Turn; also in this case the distance for Q and RANDOM does not decrease over time.

These observations point to issues associated with the 
reward function (defined by Haq \textit{et al.} and adopted unchanged in our extension study). Specifically, this reward function has two components: 1)~a continuous value determined by the inverse of the distance between the EV and the VIF, and 2)~a very large constant value ($1.0$E$+06$) when a collision occurs. 
The first component is dense, i.e., it is given to the agent at each time step, while the second one, although larger than the first one ($1.0$E$+06$ against the first one that ranges between 0 and 100), 
is sparse as it is only given when a collision occurs. As collisions are rare, in  Left-Turn and Right-Turn scenarios, the agent tends to privilege the immediate reward by minimizing the distance between the EV and the VIF. In particular, the agent, which controls the VIF, learns to steer to move backward, in order to get closer to the EV. In the Left-Turn route, this behavior does not lead to collisions, while in the Right-Turn route, where the two components of the reward function positively correlate with each other, the DQN agent learns to minimize the distance, which eventually leads to  a collision. In summary, the inverse distance used as a dense reward component 
is not always guiding the agent toward violations of the R2/DV safety requirement.

\subsubsection{Qualitative evaluation}\label{sec:qualitative_eval}

\begin{figure*}[t]
	\centering
	\includegraphics[width=0.99\textwidth]{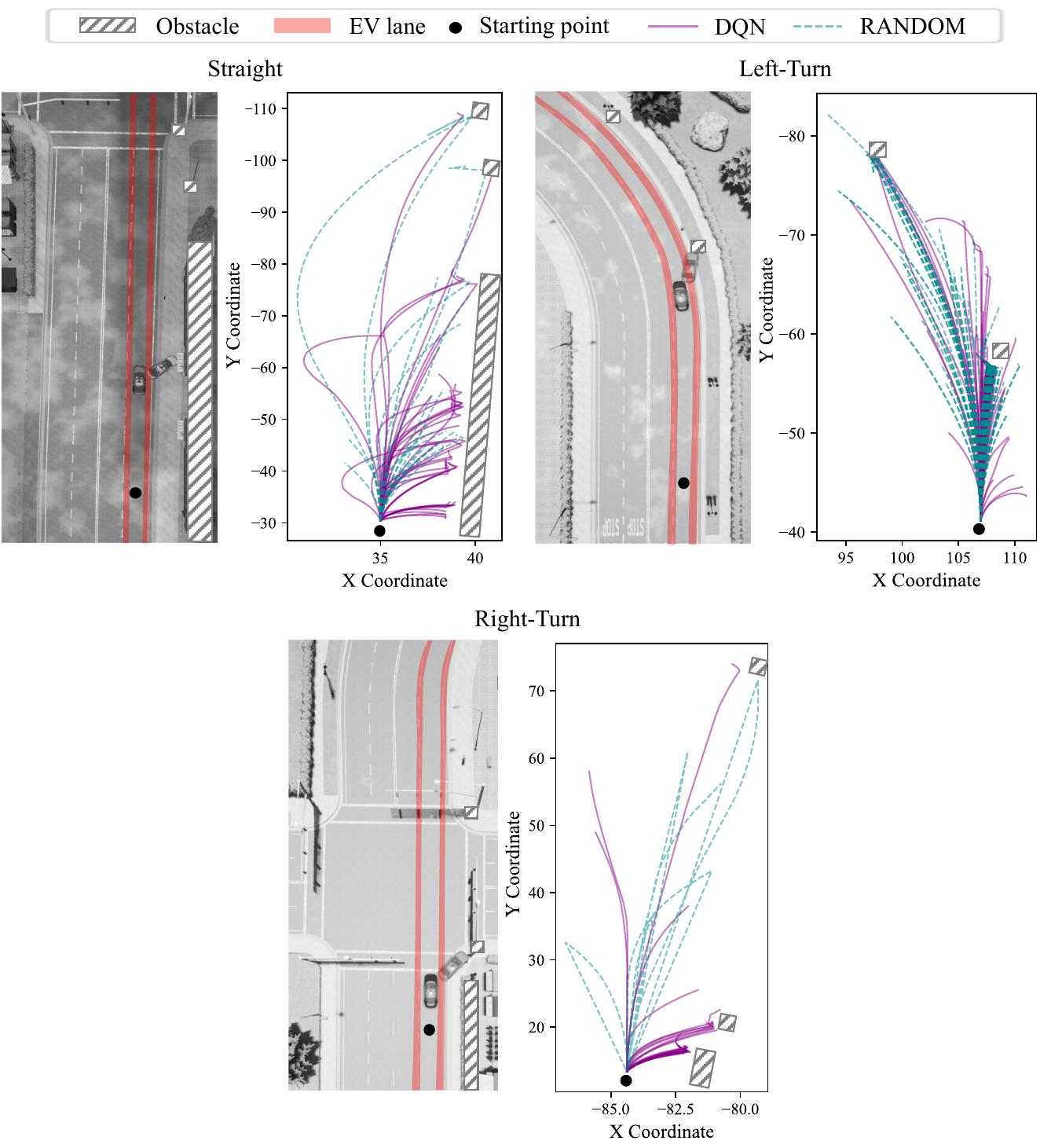}
\caption{Failure trajectories of the Vehicle in Front.} 
	\label{fig:ext_trajectories}
\end{figure*}

To analyze the different effectiveness of DQN in the three routes, we qualitatively evaluate the violations triggered by DQN and those triggered by RANDOM.
We do not show the violations of the Q agent, as  results show that Q is statistically indistinguishable from RANDOM both in terms of effectiveness and efficiency\footnote{We report the violations of the Q agent in our replication package.}.
During the execution of the testing agents, we log the $x$ and $y$ coordinates of the VIF, and we keep only the trajectories that result in a violation across all 20 repetitions. Figure~\ref{fig:ext_trajectories} shows, for each route, on the left-hand side, a bird's eye view of a failing scenario in the route, highlighting the starting point of the EV (black circle), the most relevant obstacles in the route (zebra-striped rectangles), and the lane the EV is expected to follow (delimited by two solid red lines).
On the right-hand side, we show the failure trajectories of the VIF for each technique, as well as the obstacles that are visible in the left-hand side view.

The failure trajectories show, for each route, that whenever the VIF stops and partially occupies the EV lane, the ADS under test controlling the EV is not able to avoid it, resulting in a collision (the three bird's-eye views in Figure~\ref{fig:ext_trajectories} show such collisions in each route).
This occurs in two main cases. The first one is when the sequence of VIF actions leads it to brake and stop in the critical zone. This case accounts for all the trajectories in Figure~\ref{fig:ext_trajectories} that do not terminate close to an obstacle. The second case occurs when the VIF collides with an obstacle such that the tail of the vehicle remains partially in the EV lane.

Figure~\ref{fig:ext_trajectories} shows that, independently of the route, the most common type of failure trajectory is the second one. In particular, considering the Straight and Right-Turn routes (left and right plots respectively), we observe that DQN learns to collide the VIF with the obstacles on the right to effectively find a high number of violations. The behavior of DQN, in terms of failure trajectories, is very similar in these two routes. In fact, the Right-Turn route has a long straight road before the turn, and all the failure trajectories terminate before the turn.
In these two routes, the two components of the reward function positively correlate. Indeed, the agent, by steering in one direction to minimize the distance, makes the VIF collide with obstacles and, at the same time, this maneuver causes the EV to collide with the VIF. This way, the agent gets positively rewarded at each time step by minimizing the distance (first component), and it receives a large constant reward due to the collision with the EV and the violation of the DV requirement (second component).

In the Left-Turn route (center plots), the turn is very close to the starting point, playing an active role in most of the failure trajectories. We observe that most violations found by RANDOM occur when the VIF collides with the first obstacle, located at the beginning of the curve, yet in a straight line from the starting point. Its position contributes to making the RANDOM approach effective in this route, together with the way the action space is defined. 

We identify two reasons why the formulation of the action space is critical in this route. First, by selecting random actions, there is a small chance of choosing actions that affect the VIF trajectory (only 4 out of 17 actions increase/decrease the throttle or steer left/right; the others manipulate environment variables, such as the luminosity). Since the actions affecting the VIF consist of small variations of the throttle and steering commands, an action needs to be repeatedly selected to meaningfully affect the trajectory of the VIF in the short distance that separates the starting point from the obstacle. Second, even assuming that RANDOM repeatedly selects VIF-related actions, there is the possibility of selecting counter-actions, i.e., increase the throttle when the previous action decreased it, or steer right when the previous action turned the steer to the left, resulting in the VIF going straight. Indeed, Figure~\ref{fig:ext_trajectories} shows that the trajectories of the VIF under the RANDOM agent are mostly straight, which is one of the failing conditions of the ADS under test in the Left-Turn route.

In the Left-Turn route, the DQN agent follows the immediate reward of the first component of the reward function, which privileges steering actions to minimize the distance between the EV and the VIF. This makes collisions between the two vehicles very infrequent. Correspondingly, the agent is not able to trade off immediate and long-term rewards, within the given time budget.

\subsection{Discussion}

The extension study highlights both the advantages and the limitations of using RL for testing ADS in the settings of Haq \textit{et al.}~\cite{Haq23}.  
We show that DQN is able to learn an effective policy  that   triggers significantly more violations of the ADS under test than both Q and RANDOM in most of the routes (2 out of 3). In addition, DQN is also more efficient than compared approaches in finding such violations, which is critical in the context of ADS testing where the execution of the tests is extremely time-consuming.

Overall, our extension study supports the claim that a DRL agent such as DQN, can converge to an effective policy, significantly improving effectiveness and efficiency over random testing. 
This result enables the design of novel RL-based techniques to test ADS in complex driving simulators such as CARLA. However, this requires researchers to carefully reformulate the RL problem, including the definition of states and actions spaces, as well as the reward function. Indeed, our extension study also pinpoints the limitations of the existing design. Specifically, we identified three issues in the RL formulation by Haq \textit{et al.} that need to be addressed to fully exploit the RL framework as a testing tool for ADS.

First, the two (dense and sparse) components of the reward function are not always connected with each other, making them sometimes inadequate  
to effectively guide the agent in complex scenarios such as the Left-Turn route. A possible alternative is the one proposed by Lu \textit{et al.}~\cite{Lu23}, who use the probability of collision as a reward function. %
The collision probability accounts for lateral and longitudinal distances, as well as for additional metrics such as the speed and positions of all the actors in the testing scenario. 

The second issue concerns the definition of the state space of the RL agent. Indeed, in the formulation by Haq \textit{et al.}~\cite{Haq23} the state includes absolute coordinates, which unnecessarily increases the dimensionality of the state space. For instance, two configurations of the environment that are similar in terms of relative positions of the EV and VIF vehicles but occur in different locations of the route, are encoded as different states. This reduces the possibility for the agent to reuse previously acquired knowledge, potentially increasing the training time.
One way to overcome this issue is to encode relative variables in the state space
such as ego-centric polar coordinates, or lane-centric coordinates~\cite{Leurent18}. 
This can increase the generalization capabilities of the RL agent in those situations that require the same behavior. 

The third issue is related to the definition of the action space. The actions are discrete in the setting by Haq \textit{et al.}, and they are designed to slightly perturb the dynamics of actors in the environment, such as slightly increasing/decreasing the throttle of the VIF, or making it slightly steer left/right. The effects of such small changes are delayed, as multiple small perturbations are needed to create a meaningful change. This slows down the learning process for the agent, as it needs to assign credit to the actions that lead to a high reward, despite their delayed effect. 
A possible solution to speed up learning, is to use a mixture of two common strategies. The first strategy is using an Observation Time Period (OTP)~\cite{Lu23} (also called \textit{frame skipping} in the RL literature~\cite{bellemare2012investigating,mnih2015human}), which consists of applying an action and waiting for that action to have an effect on the environment, by \textit{pausing} the RL agent for a predefined number of simulation steps (in the current setting, the RL agent acts at each simulator step). This way, the reward computed for a certain action comprises multiple simulation steps, giving more precise feedback to the agent. %
The second strategy consists of creating a layer of abstraction between the decision-making policy of the agent, and the low level actions needed to control dynamic actors in the environment (i.e., low level controls of throttle and  steering angle for the VIF)~\cite{highway-env}. In this setting, the actions available to the agent would be \textit{meta-actions}, such as ``change lane'', ``overtake'', and ``stay idle''. The low level controller, running at a higher frame rate than the RL policy, takes care of translating the meta-actions to the actual commands, leaving the RL agent the responsibility to make the most important and relevant decisions.

In summary, our extension study shows that Reinforcement Learning is a promising framework for testing Autonomous Driving Systems in simulation. Further research is needed to address more complex scenarios. This demands a thorough reformulation of the problem, which is out of the scope of this paper.

%% file: sections/related.tex
Online testing of ADS has been widely investigated, with many approaches to generate scenarios that cause the system to misbehave \cite{icse18-abdessalem,asfault,deepjanus,Majumdar19,Tuncali18,Haq22,Calo20, Klischat19,Li20}. Among the proposals, search-based techniques resulted to be particularly effective. Before MORLOT, Haq \textit{et al.}~\cite{Haq22} introduced SAMOTA, a technique that utilizes surrogate models to predict the outcome of a test case without executing it. In this context, a test case is defined as a static configuration of the environment in which the ADS operates, including factors such as the type of road and weather parameters. Riccio and Tonella~\cite{deepjanus} proposed DeepJanus to generate frontier inputs, i.e., similar input pairs that cause the ADS to misbehave for one input while working correctly for the other. Calò \textit{et al.}~\cite{Calo20} employed search-based techniques to identify \textit{avoidable} collision scenarios—scenarios where the collision would not have occurred with a reconfiguration of the ADS. They first searched for a collision and then for an alternative configuration of the ADS to avoid it.

These proposals do not deal with sequential interactions at runtime, which are required to manipulate dynamic objects in the environment (e.g., another vehicle). In this context, the use of reinforcement learning for online testing of ADSs, is gaining increasing interest. %
Besides the study by Haq \textit{et al.} \cite{Haq23}, a relevant work is the one by Lu \textit{et al.}~\cite{Lu23}, who present DeepCollision. Similarly to MORLOT, they propose an RL-based learning technique that dynamically changes the environmental conditions to find collisions with vehicles, pedestrians, and static obstacles (corresponding to violations \textit{V2}, \textit{V3}, and \textit{V4} in Section~\ref{subsec:problem_def}). DeepCollision uses DQN as RL agent to select actions from a set of 52 options to control the weather, time of the day, and the high-level behavior of actors (e.g., pedestrian crossing the road and vehicle keeping/switching lane). The state is defined by a set of 12 variables including the traffic lights color, the EV kinematics (position, speed, and rotation), and the weather conditions. The authors employ Apollo~\cite{apollo} as ADS under test and LGSVL~\cite{lgsvl} as simulator. 
However, LGSVL is unmaintained since the beginning of 2022, and the cloud servers used for deployment are no longer operational.

Other techniques use Adaptive Stress Testing (AST), a method initially employed by Lee \textit{et al.}~\cite{Lee15}, to test an aircraft collision avoidance system. AST formulates the problem of finding the most likely failure scenarios as a Markov decision process, which can be solved by RL agents. Koren \textit{et al.}~\cite{Koren18} explore the application of AST to find collisions in pedestrian crossing scenarios by extending it with deep reinforcement learning. In a short paper, Corso \textit{et al.}~\cite{Corso19} also use AST, focusing on the reward formulation to find diverse and avoidable scenarios\footnote{They consider some scenarios to be unavoidable (e.g., a pedestrian causing a collision with a stopped ADS).} as failing scenarios. 

Sharif and Marijan~\cite{Sharif22} define a multi-agent environment in which a set of deep RL agents are trained to drive adversarial cars. The goal is to find failure states of the ADS under test, defined as states in which the EV goes offroad or collides with other obstacles. The ultimate objective of the authors is to improve the ADS robustness by retraining it with adversarial inputs.

The advances in applying reinforcement learning for online testing of ADSs are undeniable. However, the proposed techniques involve highly intricate and varied simulators, defining complex RL environments that require thorough study for drawing accurate conclusions. Our work highlights the significance of replication studies in this context, which have not been conducted thus far.

%% file: sections/conclusions.tex
Scientific research on testing core software components - driven by artificial intelligence - of Autonomous Driving Systems has shown that Reinforcement Learning can be highly beneficial in supporting this difficult and time-consuming task. However, replication studies still lack in this field, despite their importance to establish well-grounded scientific evidences in empirical software engineering. 

We have replicated a recent study which showed the superiority of RL, combined with many objective search, with respect to random testing for ADSs.
While not confirming the results in the original study, the replication has provided insights on how to design RL algorithms tailored to the specific domain. 

We have thus extended the original study, showing that deep RL, with an agent designed to better fit the characteristics of the state and action spaces of the RL problem for testing ADSs, does actually outperform random testing in effectiveness and efficiency in covering their functional and safety requirements. Further empirical studies are needed to address for more complex scenarios.